# Modular and Landen transformations between finite solutions of Sine-Gordon Equation with phase **N**=**1**

Nan-Hong Kuo and C. D. Hu

Department of Physics, National Taiwan University, Taipei, Taiwan, Republic of China

Abstract

Among finite solutions of Sine-Gordon Equation with in phase **N**=**1**, we find **Modular** and **Landen transformation** between $\theta$ representation and directly integrated solutions of breather and kink. Because of these transformations, we can say we unified the finite solutions of sine-Gordon equation with phase **N**=**1**.

Key words: Modular transformation; Landen transformation; finite solutions with N=1; Sine-Gordon equation; breather; kink

# 1. Introduction:

Transformation between different structures can often be found in physics, for example, Feynmann propagator and canonical quantization, etc.. Both methods are equivalent through transformation and each has its advantage. For this reason, we study different solutions in sine-Gordon equation (**SGE**) with phase **N**=1. At first looking, they are different, but we find these solutions can be transformed to each other through **Modular** and **Landen transformations**.

In the subject of integrable systems, finite solutions (also called algebro-geometric solutions) of **SGE** can be expressed as $\theta$ functions, see Ref. [1~4]. This representation is related to Riemann surface, and has topology meanings. On the other hand, we can directly integrate out the finite solution in phase (**N**=1). So we are curious if these direct integration solution can also be expressed as $\theta$ functions. In this article, we prove there are transformations between them, that is "directly integrated solutions" in phase **N**=1 of **SGE** can be expressed as $\theta$ function. Also we find **modular** and **Landen transformations** between modular parameters of breather and kink. These transformations unify solutions of **SGE** in **N**=1.

In **Sec**.2, we give direct integration solutions of **SGE** with **N**=1. In **Sec**.3, we list $\theta$-representation solutions built from Ref. [1~4] and their spectra. In **Sec**.4, we give the method of transformation between infinite traveling wave solutions and $\theta$-representation solutions in Ref. [5]. The contents of **Sec**.5 , **6 and 7** are our main contributions. In **Sec**.5, by comparing expansion coefficients of directly integrated solutions and $\theta$-representation functions, we get their transformation. These transformations are non-trivial and interesting. We not only confirm the statement that $\theta$- representation solution are all finite solutions of **SGE** partly shown in Ref. [1~4], but also get the real form of transformation between them. There are also existing **modular and Landen transformations** between breather and kink in Appendix D of Ref. [5]. We explain the sources of these transformations by explicit calculation in **Sec**.6. In **Sec**. 7, and give the proof using transformation in **Sec**. 5. Finally, **Sec**. 8 is the conclusion.

# 2. Direct integration solutions

## 2.1 Classical solutions by direct integration

In the phase **N**=1 of **SGE**, we are only allowed one parameter, so we simply let **SGE** in this case be: $q_{tt} + \sin q = 0$. Multiplying $q_t$ to both sides and integrating over $t$, we get $\frac{1}{2}q_t^2 - \cos q = C$, where $C$ is a constant. We set $C = H - 1$, where $H$ is the conserved energy. After a short calculation, we get $\int^q \frac{dq}{\sqrt{2(H+\cos q-1)}} = \int dt = t - t_0$. We will discuss the solutions of the breather and kink separately. The subscripts $b$ and $k$ are used to denote the direct integration solutions of breathers and kinks respectively.

### Case1: Breather solution

Let $S = \sqrt{\frac{2}{H}} \sin(\frac{q}{2})$, $\int^q \frac{dq}{\sqrt{2(H+\cos q-1)}} = \int \frac{dS}{\sqrt{(1-S^2)(1-\frac{H}{2}S^2)}} = sn^{-1}(S; \sqrt{\frac{H}{2}})$, where $sn(x,k)$ is one of Jacobi elliptic function (JEF) and $k \leq 1$. For breathers, $H \leq 2$, which is consistent with our intuition about an oscillating pendulum. Above equation implies

$$q = 2\sin^{-1}[\sqrt{\frac{H}{2}} sn(t - t_0; \sqrt{\frac{H}{2}})] = 2i\ln[-ik_b sn(t - t_0; k_b) + dn(t - t_0; k_b)] \qquad (2.1)$$

where $k_b = \sqrt{\frac{H}{2}}$

Case 2: Kink solution

If we let $\tilde{S} = \sin(\frac{q}{2})$, the original integral is $\sqrt{\frac{2}{H}} \int \frac{d\tilde{S}}{\sqrt{(1-\tilde{S}^2)(1-\frac{2}{H}\tilde{S}^2)}} = \sqrt{\frac{2}{H}} sn^{-1}(\tilde{S}; \sqrt{\frac{2}{H}})$. For a kink, $H \geq 2$, which means energy is large enough to let pendulum go around in circles. It implies

$$q = 2\sin^{-1}[sn(\sqrt{\frac{H}{2}}(t-t_0); \sqrt{\frac{2}{H}}) = 2i\ln[-isn(\frac{t-t_0}{k_k}; k_k) + cn(\frac{t-t_0}{k_k}; k_k)] \quad (2.2)$$

where $k_k = \sqrt{\frac{2}{H}}$

The Reciprocal transformation of the modular parameter lead to transformation between kinks and breathers, see Ref. [9]:

$$k_b = \sqrt{\frac{H}{2}} = \frac{1}{\sqrt{\frac{2}{H}}} = \frac{1}{k_k} \Rightarrow sn(\frac{t}{k_k}; k_k) = k_b sn(t; k_b) \text{ and } cn(\frac{t}{k_k}; k_k) = dn(t; k_b) \quad (2.3)$$

By expanding the definition range of modular parameter $k$ ( p.s. traditional definition is: $-1 \leq k \leq 1$), we can say these two solutions have the same form. We can also say the breather and kink, in **eq**. (**2.1**), (**2.2**) are "equivalent" from the point of view of " **Modular transformation**", see Appendix A. **Reciprocal transformation** can expressed as combinations of two generators of **Modular transformations**.

Above is the finite solutions of **SGE**, which can be generalized to " traveling wave solution". As for the spectrum, if $H = 2 \Rightarrow k_f = \sqrt{\frac{H}{2}} = 1$, then $sn(t - t_0; k_f = 1) = \tanh(t - t_0)$, so:

$$q = 2\sin^{-1}[\tanh(t - t_0)] = 4\tan^{-1}(\frac{e^t \pm 1}{e^t \mp 1}) = 4\tan^{-1}(e^t) \pm 2\pi \quad (2.4)$$

This is the well known " traveling wave solution". We can write kink/antikink solution as the following:

$$q_{K/AK} = 4\tan^{-1}(e^{\pm\phi}) = 2i\ln(\frac{1 \mp ie^\phi}{1 \pm ie^\phi}); \phi = \frac{x - x_0 - vt}{\sqrt{1 - v^2}} \quad (2.5)$$

## 3. Θ −representation solutions

It is well known that finite solutions of soliton equations, like **SGE** or **Kdv** equation can be expressed as **Riemannian Θ function**. In **SGE**, $q_{tt} - q_{xx} + \sin q = 0$, finite solutions are

$$q = 2i\ln[\frac{\Theta(l + \frac{1}{2} + \Delta; B)}{\Theta(l + \Delta; B)}] \quad (3.1)$$

where $\Theta(l; B) \equiv \sum_{\vec{k} \in Z^N} \exp\{i\pi(< \overleftrightarrow{B}\vec{k}, \vec{k} > +2 < \vec{l}, \vec{k} >)\}$ and $\Delta$ is the Riemann's constant, which can be absorbed as the reference point $t_0$ inside $l$. $N = 1$ means one phase solution of $q_{tt} + \sin q = 0$, so we sum integer $k$ only with scalar parameters $l$ and $B$. As a result, in Θ −rep. solution

$$q = 2i\ln[\frac{\theta_4(l + \Delta; B)}{\theta_3(l + \Delta; B)}] = 2i\ln[\sqrt{k'} nd(u; k)] \quad (3.2)$$

where $\text{Im } l = at$, $a = \frac{1}{4\sqrt{k'}K}$ and $u = 2Kl$. These elliptic functions can be found in related

books, see for example Ref. [6], [7], [8]. The spectrum $\sum$ of real solutions in
$\Theta$ −representation are

Case 1, breather solution

$$\sum^{(b)} = \{E_1 = \frac{1}{16}e^{-i\varphi}, E_2 = E_1^*, \pi \leq \varphi \leq 2\pi\}; \quad \text{Re}(l) = 0; \quad \text{Re}(B) = \frac{\pm 1}{2}. \quad (3.3)$$

In section 5, we will use subscript 1 to represent the $\theta$-representation of breather.

Case 2, Kink solution

$$\sum^{(k)} = \{E_1 < E_2 < 0, E_1 = \frac{-1}{16}e^{\eta}; E_2 = \frac{-1}{16}e^{-\eta}\}; \quad \text{Re}(l) = \pm\frac{1}{4}; \quad \text{Re}(B) = 0. \quad (3.4)$$

In section 5, we will not use any subscript for the $\theta$-representation of kink.

# 4. Transformation from the $\Theta$ −representation to the traveling wave solutions

Before beginning our work, let us see how others treated the similar problem. In Ref. [5], the author use infinite-product representation of $\Theta$ functions to transform the solution into summation representation of the traveling waves. This is successful for infinite traveling wave solution, but seems failed for finite ones. For N=1, the $\Theta$-representation solution of **SGE** is

$$q = 2i\ln[\frac{\Theta(l+\frac{1}{2};B)}{\Theta(l;B)}] = 2i\ln[\frac{\theta_4(l;B)}{\theta_3(l;B)}] = 2i\ln[\frac{i\theta_2(l+\frac{B}{2}+\frac{1}{2};B)}{\theta_2(l+\frac{B}{2};B)}] \quad (4.1)$$

In Ref. [9], we can find infinite product representation of $\theta_2(l;B)$:

$$\theta_2(l;B) = c \cdot e^{\pi i l} \cdot \prod_{n=1}^{\infty}[1+e^{2\pi i(nB-l)}] \cdot \prod_{n=-\infty}^{0}[1+e^{-2\pi i(nB-l)}]. \quad (4.2)$$

Here we choose $l = \frac{1}{4} + \frac{i}{2\pi}[\kappa(x-x_0)+wt]$, and $\kappa$ and $w$ come from $\theta$-representation calculus in (V.16) of ref. [9]. So

$$\frac{\theta_4(l;B)}{\theta_3(l;B)} = \prod_{n=1}^{\infty}[\frac{1-ie^{\alpha_n}}{1+ie^{\alpha_n}}] \cdot \prod_{n=-\infty}^{0}(-1) \cdot [\frac{1-ie^{\alpha_n}}{1+ie^{\alpha_n}}] \quad (4.3)$$

where $\alpha_n = \kappa(x-x_0)+wt+2n\pi iB$. In the kink case, Re$[B] = 0$, so

$$q = \sum_{n=-\infty}^{\infty}\{2i\ln[\frac{1-ie^{\theta_n}}{1+ie^{\theta_n}}]+\pi(sgn(n)-1) \quad (4.4)$$

$$= \sum_{n=-\infty}^{\infty}\{q_K(x-x_0-nL)+\pi(sgn(n)-1)$$

where $sgn = \begin{cases} +1, \text{ for } n>0 \\ -1, \text{ for } n\leq 0 \end{cases}$, and $q_k$ can be found in eq. (2.5).

In the breather case, Re$[B] = \frac{1}{2}$, the solution is

$$q = \sum_{n=-\infty}^{\infty}\{2i\ln[\frac{1-ie^{\theta_{2n}}}{1+ie^{\theta_{2n}}}]+2i\ln[\frac{1+ie^{\theta_{2n-1}}}{1-ie^{\theta_{2n-1}}}]+2\pi(sgn(n)-1)\} \quad (4.5)$$

$$= \sum_{n=-\infty}^{\infty}\{q_K(x-x_0-2nL)+q_{AK}(x-x_0-(2n-1)L)+2\pi(sgn(n)-1)\}$$

where the sum of $q_K$ and $q_{AK}$ gives the breather solution. So from the $\theta$- representation solution, we can obtain " kink trains" and " breather trains". This is the technique using infinite product of $\theta$ function list in Ref. [5].

# 5. Transform between direct integration solutions and $\theta$-representation solutions

In this section, we obtain the transformation between directly integrated solutions and $\theta$-representation solutions by expanding the **Jacobi elliptic function** (**JEF**) in series. In Sec. 7, we use these transformation to prove our statement completely.

## (A) Breather:

We are going to use the following formulas in our proof. These will be useful to build connection between eq. (2.1) and the $\theta$-representation solution in eq. (3.1). From Ref. [9], if $\text{Im}\frac{u}{K} < \text{Im}\frac{iK'}{K}$, we have following expansion:

$$sn(u;k) = \frac{2\pi}{kK}\sum_{m=0}^{\infty}\frac{q^{m+\frac{1}{2}}}{1-q^{2m+1}}\sin[(m+\frac{1}{2})\frac{\pi u}{K}], \tag{5.1}$$

$$dn(u;k) = \frac{\pi}{2K} + \frac{2\pi}{K}\sum_{m=0}^{\infty}\frac{q^{m+1}}{1+q^{2m+2}}\cos[(m+1)\frac{\pi u}{K}],$$

$$nd(u;k) = \frac{\pi}{2k'K} + \frac{2\pi}{k'K}\sum_{m=0}^{\infty}(-1)^{m+1}\frac{q^{m+1}}{1+q^{2m+2}}\cos[(m+1)\frac{\pi u}{K}].$$

In N=1 breather, $\overleftrightarrow{B}$ in eq. (3.1) is equal to $\tau_1 = \frac{iK'_1}{K_1}$. And in eq. (5.1), we must use the standard notation, $q_1 = e^{\pi i\tau_1} = e^{-\pi\frac{K'_1}{K_1}}$. In another way, by the condition of eq. (3.3), we require $\text{Re}(\tau_1) = \frac{1}{2}$, so here $K_1$ and $K'_1$ are not real as they are in the usual cases. As a result, $q_1 = e^{\pi i(\frac{1}{2}+\frac{i\widetilde{K}'}{\widetilde{K}})} = i\widetilde{q}$ where $\widetilde{q} = e^{\pi i\widetilde{\tau}} = e^{-\pi\frac{\widetilde{K}'}{\widetilde{K}}}$. The other condition of a breather is $l = iat \Rightarrow u = 2Kl = i2Kat$. and $a$ is a constant. So in $nd(u;k)$, the $\Theta$–rep. solution, we will have real/Imaginary part when $m$ is odd/even. And $k'_1$ and $K_1$ are parameters of following breather $\Theta$–rep. solution, the argument of $2i\ln$ of $\theta$-representation of breather in eq. (3.2)

$$\sqrt{k'_1}\,nd(2iK_1at;k_1) = \frac{1}{\sqrt{k'_1}}\{[\frac{\pi}{2K_1} + \frac{2\pi}{K_1}\sum_{m:odd}\frac{q_1^{m+1}}{1+q_1^{2(m+1)}}\cos[(m+1)\frac{\pi u}{K_1}] \tag{5.2}$$

$$- [\frac{2\pi}{K_1}\sum_{m:even}\frac{q_1^{m+1}}{1+q_1^{2(m+1)}}\cos[(m+1)\frac{\pi u}{K_1}]]\},$$

$$\text{Re}[\sqrt{k'_1}\,nd(2iK_1at;k_1)] = \frac{1}{\sqrt{k'_1}}[\frac{\pi}{2K_1} + \frac{2\pi}{K_1}\sum_{n_1=0}^{\infty}\frac{(-1)^{n_1+1}\widetilde{q}^{2n_1+2}}{1+\widetilde{q}^{4n_1+4}}\cos[(2n_1+2)\cdot 2\pi iat] \tag{5.3}$$

from $m = 2n_1 + 1$, an odd number, and

$$\text{Im}[\sqrt{k'_1}\,nd(2iK_1at;k_1)] = \frac{-1}{\sqrt{k'_1}}[\frac{2\pi}{K_1}\sum_{n_2=0}^{\infty}\frac{(-1)^{n_2}\widetilde{q}^{2n_2+1}}{1-\widetilde{q}^{4n_2+2}}\cos[(2n_2+1)\cdot 2\pi iat] \tag{5.4}$$

from $m = 2n_2$, an even number.

$2K_1a = \frac{1}{2\sqrt{k'_1}}$ because we require $2i\ln[\sqrt{k'_1}\,nd(2iK_1at;k_1)]$ is a solution of $u_{tt} + \sin u = 0$, the **SGE** with N=1. But this relation is right only in normal range of $k'_1$, that is $0 \leq k'_1 \leq 1$. Later this relation will change because the range of $k'_1$ is expanded by transformation.

See also Appendix A for modular transformation of $k'_1$.

On the other hand, the direct integration solution is $2i\ln[dn(t - t_0; k_b) - ik_b sn(t - t_0; k_b)]$ in eq. (2.1). The real part is

$$dn(t - t_0; k_b) = \frac{\pi}{2K_b} + \frac{2\pi}{K_b} \sum_{m=0}^{\infty} \frac{q_b^{m+1}}{1 + q_b^{2(m+1)}} \cos[(m+1)\frac{\pi(t - t_0)}{K_b}] \qquad (5.5)$$

where $q_b = e^{\pi i \tau_b}$, $\tau_b = \frac{iK'_b}{K_b}$ is the periodic matrix of direct integration solution. The imaginary part is

$$-k_b sn(t - t_0; k_b) = \frac{-2\pi}{K_b} \sum_{m=0}^{\infty} \frac{q_b^{m+\frac{1}{2}}}{1 - q_b^{2m+1}} \sin[(m + \frac{1}{2})\frac{\pi(t - t_0)}{K_b}]. \qquad (5.6)$$

Comparing real and imagery parts in eq. (5.3)~(5.6), we obtain following relations:

$$(1)\ q_b = \tilde{q}^2;\ (2)\ t_0 = K_b;\ (3)\ K_b = \sqrt{k'_1}\, K_1 \qquad (5.7)$$

and also

$$4ia = \frac{1}{K_b} = \frac{1}{\sqrt{k'_1}\, K_1}. \qquad (5.8)$$

This relation is a little different from before ( i.e. $2K_1 a = \frac{1}{2\sqrt{k'_1}}$) because if we transform from direct integration breather into $\theta$-representation structure, $k'_1$ become negative!

Relation (1) is the "**Landen transformation**" ( hence the period is doubled, $\tau_b = 2\tilde{\tau}$). Here we give some details of the proof of these relations. Beginning from $\tau_1$ in Appendix D of Ref. [5]: $\tau_1 = \frac{1}{2} + \frac{iK(s'_b)}{2K(s_b)}$, $s_b = \cos(\frac{phE_2}{2})$, $s'_b = \sin(\frac{phE_2}{2})$. we found that $\tau_1$ can be calculated from loop integration of spectrum in **eq**. (3.3). Let $\tau_b = \frac{iK(s'_b)}{K(s_b)}$, then we can see $\tau_1 = \frac{1+\tau_b}{2}$; where $\tau_b$ is the product of the combination of modular and Landen transformation. ( i.e. $\tau_b \to \tau_2 = 1 + \tau_b \to \tau_1 = \frac{\tau_2}{2} = \frac{1+\tau_b}{2}$). So $K_b = k'_2 K_2 = \frac{2\sqrt{k'_1}}{1+k'_1} K_2 = \sqrt{k'_1}(1 + k_2)K_2 = \sqrt{k'_1}\, K_1$. Here we use relations list in **Ref**. [**9**,**10**]. This is relation (3) of **eq**. (**5.7**). Since $q_b = -q_2 = -q_1^2 = \tilde{q}^2$, we get relation (1) of **eq**. (**5.7**). Here we use $\tau_1 = \frac{1+\tau_b}{2}$ to prove **eq**. (**5.7**). In **Section 6**, we will focus on breather relation: $\tau_1 = \frac{1}{2} + \frac{iK(s')}{2K(s)}$ and kink relation.

## (B) Kink:

As for the case of kinks, we will use another expansion as the following. These formulas will be used to show the equivalence between the form in eq. (2.2) and the kink $\theta$-representation in eq. (3.1). From Ref. [9], if $\text{Im}\,\frac{u}{K} < \text{Im}\,\frac{iK'}{K}$, we have another expansion:

$$sn(u; k) = \frac{\pi}{2kK} \sum_{m=-\infty}^{\infty} \csc \frac{\pi}{2K}[u - (2m - 1)iK'] \qquad (5.9)$$

$$cn(u; k) = \frac{\pi i}{2kK} \sum_{m=-\infty}^{\infty} (-1)^m \csc \frac{\pi}{2K}[u - (2m - 1)iK']$$

In order to make comparison, we expand the argument of $2i\ln$ of the directly integrated solution in eq. (2.2) with csc functions as the following:

$$-isn(\frac{t}{k_k};k_k) + cn(\frac{t}{k_k};k_k) = \frac{-\pi i}{k_k \cdot K_k} \sum_{m:odd} \csc \frac{\pi}{2K_k}[\frac{t}{k_k} - (2m-1)iK'_k] \tag{5.10}$$

$$= \frac{-\pi i}{k_k \cdot K_k} \sum_{m'=-\infty}^{\infty} \csc[\frac{\pi}{2K_k}\frac{t}{k_k} + (\frac{3}{2} - 2m')\frac{\pi i K'_k}{K_k}]$$

where $m = 2m' - 1$. From $\theta$-representation solution in eq. (3.2), and the relations (1) $sn(iv;k) = isc(v;k')$, (2) $\frac{1}{dn(v+iK';k)} = isc(v;k)$, the argument of $2i\ln$ is

$$\frac{\sqrt{k'}}{dn(u;k)} = i\sqrt{k'}\, sc(u - iK';k) = \sqrt{k'}\, sn(iu + K';k') \tag{5.11}$$

$$= \frac{\pi}{2\sqrt{k'}\,K'} \sum_{m=-\infty}^{\infty} \csc \frac{\pi}{2K'}[iu + K' - (2m-1)iK].$$

Let $u = \frac{K}{2} + i2Ka(t - t_0)$, we obtain

$$\frac{\sqrt{k'}}{dn[\frac{K}{2} + i2Ka(t-t_0);k]} = \frac{\pi}{2\sqrt{k'}\,K'} \sum_{m=-\infty}^{\infty} \csc[\frac{\pi}{2} - \frac{\pi K}{K'}a(t - t_0) + (\frac{3}{4} - m)\frac{\pi i K}{K'}]. \tag{5.12}$$

Comparing the over-all coefficients outside summations and each coefficient of $(t - t_0)$ in eq.(5.10) and (5.12), we obtain the following relations

$$(1)\ 2\sqrt{k'}\,K' = ik_k K_k, \tag{5.13}$$
$$(2)\ \frac{1}{2K_k k_k} = \frac{-Ka}{K'}.$$

We can let $\tilde{\tau}_k \equiv \tau_k - 1$ ( mod 2), which is equivalent to adding $\frac{-3\pi}{2} = \frac{\pi}{2}$ ( mod $2\pi$) to the argument of csc function. (This is one of the "**Modular transformation**", see Appendix A.) So eq.(5.10) and (5.12) are identical. Finally, we obtain the transformation between them

$$-\frac{1}{\tau} = 2(\tau_k - 1) \tag{5.14}$$

where $\tau = \frac{iK'}{K} \Rightarrow -\frac{1}{\tau} = \frac{iK}{K'}$; $\tau_k = \frac{iK'_k}{K_k}$.

$\tau \to 2\tau$ is "**Landen transformation**" appear in eq. (5.14). This is equivalent to $k_1 = \frac{1-k'}{1+k'}$ in **JEF**, where $k_1 = \frac{[\theta_2(0;2\tau)]^2}{[\theta_3(0;2\tau)]^2}$ and $k = \frac{[\theta_2(0;\tau)]^2}{[\theta_3(0;\tau)]^2}$, see Ref. [9]. Note that $\tau \leftrightarrow \frac{-1}{\tau}$ means $k \leftrightarrow k'$ and $\tau_k \to \tau_k - 1$ means $k \to \tilde{k} = \frac{ik_k}{k'_k}$ and $k' \to \tilde{k'} = \frac{1}{k'_k}$, where $\tilde{k}$ is the modulus with period $\tau_k - 1$. The relation in eq. (5.14) combines these three transformations. In parameters of **JEF**, eq. (5.14) is equivalent to the following transformation between direct integration solutions and $\theta$- representation solutions

$$k' = \frac{k'_k - 1}{k'_k + 1} \tag{5.15}$$

From eq. (5.15) we can prove directly the relation (1) of eq. (5.13). The relation (2) of eq. (5.13) is just to determine $a$, and this is consistent with eq. (5.8). Further, we get more concise relations between their periods

$$K(k') = \frac{1 + k'_k}{2} K(k_k) \tag{5.16}$$

# 6. Transformation between kink and breather in N=1

In this section, we give the derivation of the transformation between the kinks and breathers. In Ref. [6~8], the conserved energy $H = 1 - 8(E_1 + E_2) = 1 - \cos\varphi \leq 2$ from the spectrum of breather $\sum^{(b)}$ and $H = 1 + \cosh\eta \geq 2$ from that of kink $\sum^{(k)}$. So it is reasonable to let $\varphi \to \pi + i\eta$ when solution of **SGE** change from breather to kink, just like traveling wave would become evanescent wave with a phase shift when free particle meets a potential barrier higher than its total energy.

As in Table 2, Appendix D of Ref. [5], we reform two holomorphic differential integrals:

$$I(a) = \oint_a dI; \quad I(b) = \oint_b dI$$

where $dI = \frac{dE}{R(E)}$, $R^2(E) = E(E - E_1)(E - E_2)$ and $a$ and $b$ cycles are two independent loop in Ref. [5] and get the period $\tau_1 = \frac{1}{2} + \frac{iK(s'_b)}{2K(s_b)}$; ( i.e. the same notations below eq. (5.8). ), $s_b = \cos(\frac{phE_2}{2}) = \cos(\frac{\varphi}{2})$; $s'_b = \sin(\frac{\varphi}{2})$ in breather case, and $\tau_k = \frac{iK(s'_k)}{K(s_k)}$; $s'_k = e^{-\eta}$ in kink case. How to transform between them? Following is a short explanation.

## 6.1 Reciprocal and inverse modular transformation

We use the relation of the solution found by direct integration:
$k_k = \sqrt{\frac{2}{H}} \underset{(1)}{=} \frac{2}{e^{\frac{\eta}{2}} + e^{-\frac{\eta}{2}}} = \frac{1}{\cosh\frac{\eta}{2}}$, (1) is by $H = 1 + \cosh\eta$. Another direction is
$k_k \underset{(2)}{=} \frac{1}{k_b} \underset{(3)}{=} \frac{1}{\sin\frac{\varphi}{2}} \underset{(4)}{=} \frac{1}{\sin(\frac{\pi+i\eta}{2})} = \frac{1}{\cosh\frac{\eta}{2}}$, the same as above result. (2) is the **reciprocal transformation**; (3) means $k_b = s'_b$ by doing " **inverse modular transformation**" in $\tau_b = \tau_{s_b} \to \frac{-1}{\tau_{k_b}}$. (4) is $\varphi \to \pi + i\eta$. In the following we will find relations between $s'_k$ and $k_k$.

## 6.2 Landen transformation

We can also use transformation eq. (5.15), i.e. combining three major kinds of transformation mentioned there with the parameter $\frac{k'_k - 1}{k'_k + 1} = e^{-\eta} = s'$, to get $s'$ in the same form as that in Appendix D of Ref. [5]. We begin with the relation below eq. (5.8):
$\tau_1 = \frac{1}{2} + \frac{iK(s'_b)}{2K(s_b)} = \frac{\tau_2}{2} = \frac{1+\tau_b}{2}$; $s_b = \cos(\frac{phE_1}{2})$, $\tau_2 = \frac{iK(s'_2)}{K(s_2)}$, $\tau_1 = \frac{iK(s'_1)}{K(s_1)}$ and $\tau_b = \frac{iK(s'_b)}{K(s_b)}$. By modular and Landen transformation, we get $s_2 = \frac{is_b}{s'_b} = i\cot\frac{\varphi}{2}$;
$s'_2 = \csc\frac{\varphi}{2} \Rightarrow s'_1 = \frac{1-s_2}{1+s_2} = -e^{i\varphi}$, here we use "inverse Landen transformation" because of $\tau_1 = \frac{\tau_2}{2}$. Finally, if we let $\varphi \to \pi + i\eta$ as before, then $s'_1 = e^{-\eta} = s'$. So $\tau_1 = \tau = \frac{1+\tau_b}{2}$. This means we can neglect the different forms of periodic matrix in breather and kink. They are the same thing if we remember $\varphi \leftrightarrow \pi + i\eta$.

## 6.3 The reason of different form of periodic matrices between breather and kink:

Here, we want to explain why we list different periodic matrices $\tau_b$ and $\tau_k$ for breather and kink even though we have proved they are actually the same thing in Sec.6.2. If $I(a)$ and $I(b)$ are two independent holomorphic differential integrals, then in order to calculate $\tau = \frac{I(b)}{I(a)}$, the periodic matrix in Appendix D in Ref. [5], we have to use different transformations for kinks and breathers because of different spectra, eq. (3.3) and eq. (3.4). These including **Landen** and **modular transformation**. Below we calculate $I(a)$ of kinks and breathers to give an example.

### 6.3.1 $I(a)$ of kink

The spectrum is in eq. (3.4). Here we let $\widetilde{E} = \frac{E - E_1}{E_2 - E_1}$ and $t^2 = \widetilde{E}$

$$I(a) = -2 \int_{E_1}^{E_2} \frac{dE}{\sqrt{E(E-E_1)(E-E_2)}} = \frac{-2}{\sqrt{-E_1}} \int_0^1 \frac{d\tilde{E}}{\sqrt{\tilde{E}(1-\tilde{E})[1-(\frac{E_2-E_1}{-E_1})\tilde{E}]}} \quad (6.1)$$

$$= \frac{-4}{\sqrt{-E_1}} \int_0^1 \frac{dt}{\sqrt{(1-t^2)[1-(\frac{E_2-E_1}{-E_1})t^2]}} = \frac{-4}{\sqrt{-E_1}} K(k_k = \sqrt{\frac{E_2-E_1}{-E_1}})$$

where $k_k = \sqrt{\lambda_k}$ with $\lambda_k = \frac{E_2-E_1}{-E_1} \Rightarrow k'_k = \sqrt{1-\lambda_k} = e^{-\eta}$.

6.3.2 $I(a)$ of breather

The spectrum is in eq. (3.3).

$$I(a) = 4\operatorname{Re} \int_{\operatorname{Re} E_1}^{E_1} \frac{dE}{\sqrt{E(E-E_1)(E-E_2)}}. \quad (6.2)$$

Here we do "**Möbius transformation**" to change variable $E$ into $\varsigma$ with modular parameter $k_3$:

$$\frac{E-E_1}{E-E_1^*} \frac{-E_1^*}{-E_1} = \frac{\varsigma-1}{\varsigma+1} \frac{1-k_3}{1+k_3}. \quad (6.3)$$

Defining the cross ratio

$$\lambda_b = \frac{E_1^*}{E_1} = e^{2i\varphi} \quad (6.4)$$

and the relation between $\lambda_b$ and $k_3$ is

$$k_3 = \frac{1+\sqrt{\lambda}}{1-\sqrt{\lambda}} = i\cot\frac{\varphi}{2}, \quad (6.5)$$

eq. (6.2) becomes

$$I(a) = \frac{4\operatorname{Re}}{\sin\frac{\varphi}{2}\sqrt{|E_1|}} \int_k^1 \frac{-d\varsigma}{\sqrt{(1-\varsigma^2)(1-k_3^2\varsigma^2)}}. \quad (6.6)$$

Let $t = \sqrt{1-k_3^2\varsigma^2}$; $h^2 = \frac{1}{1-k_3^2} = \sin^2\frac{\varphi}{2} \Rightarrow h' = \cos\frac{\varphi}{2}$, eq. (6.6) is transformed to

$$I(a) = \frac{-4i}{\sqrt{|E_1|}} \int_1^{\frac{1}{h}} \frac{dt}{\sqrt{(1-t^2)(1-h^2t^2)}} = \frac{-4}{\sqrt{|E_1|}} K'(h) = \frac{-4}{\sqrt{|E_1|}} K(h') = \frac{-4}{\sqrt{|E_1|}} K(\cos\frac{\varphi}{2}). \quad (6.7)$$

So there are different transformations of variables and modular parameters in breather and kink.

By $h = \frac{1}{k'_3}$, and let $k_b = h' = \cos\frac{\varphi}{2} = \frac{ik_3}{k'_3}$, there are "translational **modular transformation**" between $k_b$ and $k_3$, that is $\tau_b = \tau_3 - 1$. Since in eq. (6.4), $\lambda_b = e^{2i\varphi}$, and by our rule, when breather become kink, $\varphi \to \pi + i\eta \Rightarrow \lambda_b = e^{-2\eta}$, so $k'_k = \sqrt{1-\lambda_k} = \pm\sqrt{\lambda_b}$ ( choosing minus sign). From eq. (6.5), $k_3 = \frac{1+\sqrt{\lambda_b}}{1-\sqrt{\lambda_b}} = \frac{1-k'_k}{1+k'_k}$. This satisfies "**Landen transformation**". So $\tau_3 = 2\tau_k = \tau_b + 1 \Rightarrow \tau_k = \frac{1+\tau_b}{2}$, which is consistent with the conclusion in Section 6.2. This finishes the proof that the different forms of $\tau_k$ and $\tau_b$ come from different change of variables, and this is due to different spectra.

## 7. Complete and short proof of the transformations

In Sec. 5, we expand solutions in eq. (2.2) and eq. (3.2) by series and show they are identical. In Sec. 6, we use these relations to check the formulas list in papers, like Ref. [5]. In this section, we will give another direct proof to show the transformation is correct using the identity in (3.9.24) of Ref. [10]. If $k_1 = \frac{1-k'}{1+k'}$ and $u_1 = \frac{2}{1+k_1}u$, the identity is eq. (7.1)

$$nd(u;k) = \frac{1}{1-k_1}[dn(u_1,k_1) - k_1 \cdot cn(u_1,k_1)]. \tag{7.1}$$

Using following relations

$$cn(u + K + iK', k) = \frac{-ik'}{k}nc(u,k) \tag{7.2}$$

$$dn(u + K + iK', k) = ik'sc(u,k)$$

and in view in of eq. (5.15) we get

$$k_1 = \frac{1}{k'_k} = \frac{1-k'}{1+k'}. \tag{7.3}$$

The Reciprocal Modular transformation, see Sec. 162 of Ref. [9], gives us

$$dn(u_1, k_1) = cn(k_1 u_1, k'_k) \tag{7.4}$$

$$cn(u_1, k_1) = dn(k_1 u_1, k'_k)$$

and hence,

$$dn(u_1, k_1) - k_1 \cdot cn(u_1, k_1) \tag{7.5}$$

$$= cn(k_1 u_1, k'_k) - \frac{1}{k'_k}dn(k_1 u_1, k'_k).$$

Using above relations, eq. (7.1) become the following identity:

$$(1 - k_1) \cdot nd[\frac{1+k_1}{2}(u_1 + \frac{K'_k}{k_1} + \frac{iK_k}{k_1}), k] \tag{7.6}$$

$$= cn(k_1 u_1 + K'_k + iK_k, k'_k) - \frac{1}{k'_k}dn(k_1 u_1 + K'_k + iK_k, k'_k)$$

$$= -\frac{ik_k}{k'_k}nc(k_1 u_1, k'_k) - \frac{ik_k}{k'_k}sc(k_1 u_1, k'_k)$$

$$= -\frac{ik_k}{k'_k}cn(ik_1 u_1, k_k) - \frac{k_k}{k'_k}sn(ik_1 u_1, k_k).$$

In our direct kink solutions, eq. (2.2)

$$-isn(\frac{t}{k_k}, k_k) + cn(\frac{t}{k_k}, k_k) \tag{7.7}$$

$$= i[-sn(\frac{t}{k_k}, k_k) - icn(\frac{t}{k_k}, k_k)]$$

$$= i\frac{k'_k}{k_k}(1 - k_1) \cdot nd[\frac{1+k_1}{2}(\frac{1}{ik_1}\frac{t}{k_k} + \frac{K'_k}{k_1} + \frac{iK_k}{k_1}), k].$$

Using eq. (7.3), the total coefficient outside $nd$ function is

$$(1) : i\frac{k'_k}{k_k}(1 - k_1) = \sqrt{k'}. \tag{7.8}$$

We also get coefficients inside $nd$ function in **eq. (7.7)**. Using **Reciprocal transformation**, we get

$$K(\frac{1}{k}) = k[K(k) + iK'(k)]. \tag{7.9}$$

In our case, by **eq. (7.3)**,

$$K(k_1) = k'_k[K'(k_k) + iK(k_k)] = \frac{K'_k + iK_k}{k_1} \tag{7.10}$$

with **Landen transformation** in eq. (7.3), we obtain

$$(2): \frac{1+k_1}{2}K(k_1) = \frac{K(k)}{2}. \tag{7.11}$$

This is the constant inside $nd$ function in eq. (7.7). This corresponds to the condition of kink in eq. (3.2), $\mathrm{Re}(l) = \frac{1}{4}$. Finally, the coefficient of $t$ is

$$(3): -i\frac{1+k_1}{2k_1}\frac{1}{k_k} = \frac{1}{2\sqrt{k'}}. \tag{7.12}$$

To be consistent with eq. (3.2), we must let $k' < 0$ so that the coefficient of $t$ is purely imaginary. This is consistent with eq. (7.3) because $k_k = \sqrt{\frac{2}{H}}$. It satisfies $0 < k_k \le 1$ and also $k'_k = \frac{1+k'}{1-k'}$, so $k' < 0$ is consistent with $k_k \le 1$.

Substituting $(1), (2)$ and $(3)$ into eq. (7.7), we obtain

$$\sqrt{k'}\,nd(\frac{t}{2\sqrt{k'}} + \frac{K}{2}, k). \tag{7.13}$$

The breather solution eq. (2.1) is related to $\theta$- representation by kink relation as below,

$$\begin{aligned}
&-ik_b sn(t-t_0, k_b) + dn(t-t_0, k_b) \\
&= -isn(\frac{t-t_0}{k_k}, k_k) + cn(\frac{t-t_0}{k_k}, k_k) \\
&= \sqrt{k'}\,nd(\frac{t-t_0}{2\sqrt{k'}} + \frac{K}{2}, k) = \sqrt{k'}\,nd(\frac{t}{2\sqrt{k'}}, k).
\end{aligned} \tag{7.14}$$

In the last step, we use eq. (5.7), $t_0 = K_b = \sqrt{k'}K$. From eq. (7.3) and (2) in Section 6.1, we find $k_k = \frac{1}{k'_1} = \frac{1}{k_b} \Rightarrow k_b = k'_1 = \frac{2\sqrt{k'}}{1+k'}$.

## 8. Conclusions

The transformation in Sec. 4, also mentioned in Ref. [5] is educative because the authors use infinitely product of $\theta$- function to find the relation between $\theta$-representation solution and traveling wave solutions. But we could not use this technique to transform classical pendulum solutions and $\theta$- representation solutions because we could not find "another" infinite product of classical pendulum. In Sec. 5, we expand **JEF** by different functions in breather and kink case and find the transformation. In Sec. 6, we further deepen the meaning of transformation and also find the transformation of modular parameters between kink and breather. Finally. in Sec. 7, we give direct proof confirming our transformations in Sec.5. We think finding these transformation also has application in unifying solutions in different forms.

## Appendix A: Modular transformation and equivalent class

In Weierstrass elliptic function:

$$Y^2 = 4X^3 - g_2 X - g_3 = 4(X-e_1)(X-e_2)(X-e_3). \qquad \text{A.1}$$

For convenience, we define two periods as:

$$2w_3 = \oint_{\alpha_1} \frac{dX}{Y} = 2\int_{e_1}^{e_2} \frac{dX}{\sqrt{(X-e_1)(X-e_2)(X-e_3)}} \qquad \text{A.2}$$

$$2w_1 = \oint_{\alpha_2} \frac{dX}{Y} = 2\int_{e_3}^{e_2} \frac{dX}{\sqrt{(X-e_1)(X-e_2)(X-e_3)}}.$$

where $\alpha_1$ and $\alpha_2$ are the integral loops surrounding $e_1$ and $e_2$ and $e_3$ and $e_2$.

Also define $\tau$ as the ratio of two period:

$$\tau = \frac{w_3}{w_1}. \qquad \text{A.3}$$

Since we can permute $e_1, e_2, e_3$, we have $3! = 6$ " equivalent definition of $\tau$". From Jacobi elliptic function, we can relate parameters $k, k'$ with $e_1, e_2, e_3$ as

$$k^2 = \frac{[\theta_2(0;\tau)]^4}{[\theta_3(0;\tau)]^4} = \frac{e_2 - e_3}{e_1 - e_3} \qquad \text{A.4}$$

$$k'^2 = \frac{[\theta_4(0;\tau)]^4}{[\theta_3(0;\tau)]^4} = \frac{e_1 - e_2}{e_1 - e_3}$$

**Case 1**: Choose one set of $e_1, e_2, e_3$. So $\tilde{\tau} = \tau \Rightarrow \tilde{k} = k; \tilde{k'} = k'$.

**Case 2**: $e_3 \leftrightarrow e_2$ Permuting $e_3$ and $e_2$ in the definition of $\tau$, eq. (A.3), we get $\tilde{\tau} = 1 - \tau \Rightarrow \tilde{k} = \frac{ik}{k'}; \tilde{k'} = \frac{1}{k'}$.

**Case 3**: $e_1 \leftrightarrow e_2$; then $\tilde{\tau} = \frac{-\tau}{1-\tau} \Rightarrow \tilde{k} = \frac{1}{k}; \tilde{k'} = \frac{ik'}{k}$.

**Case 4**: $e_1 \leftrightarrow e_3$; then $\tilde{\tau} = \frac{1}{\tau} \Rightarrow \tilde{k} = k'; \tilde{k'} = k$.

**Case 5**: $\tilde{\tau} = \frac{1}{1-\tau}$; It combine **Case 2** and **Case 4**; so $\tilde{k} = \frac{1}{k'}; \tilde{k'} = \frac{ik}{k'}$.

**Case 6**: $\tilde{\tau} = -\frac{1-\tau}{\tau}$; it combine **Case 3** and **Case 4**; so $\tilde{k} = \frac{ik'}{k}; \tilde{k'} = \frac{1}{k}$.

These are 6 equivalent class of Modular transformation ( also called " linear fractional transformation). The generator are **Case 2** and **Case 4**. Modular transformation is

$\tilde{\tau} = \frac{a\tau+b}{c\tau+d}$; $\begin{pmatrix} a & b \\ c & d \end{pmatrix} \in SL(2, Z)$.

# References:

[**1**]. E. D. Belokolos, et. al.,"Algebro-Geometric Approach to Nonlinear Integrable Equations", Springer-Verlag (1994).

[**2**]. F. Gesztesy, et. al., "Soliton Equations and their algebro-geometric solutions" Vol.1, Cambridge (2003).

[**3**]. J. Zagrodzinski, J. Math. Phys. **24**, 46 (1983).

[**4**]. J. Zagrodzinski; Lettere Al Nuovo Cimento **30**, 226 (1981).

[**5**]. M. G. Forest, et. al., J. Math. phys. **23**, 1248 (1982).

[**6**]. N. M. Ercolani, et. al., Comm. Math. Phys. **99**,1 (1985).

[**7**]. R.Flesch, et. al., Physica D **48** 169 (1991).

[**8**]. M.G. Forest, et. al., SIAM J. Appl. Math. 52, 746 (1992).

[**9**]. Paul F. Byrd, et. al., "Handbook of Elliptic Integrals for Engineers and Scientists", 2nd ed. Springer-Verlag (1971).

[**10**]. D. F. Lawden, "Elliptic functions and applications", Springer-Verlag (1989).